\documentclass[12pt,preprint]{aastex}

\def\Epkmax{E_{\rm  0}}
\def\Ep{E_{\rm  p}}

\def\F0{F_{\rm  0}}
\def\t0{t_{\rm  0}}

\def\P0{\Phi_{\rm  0}}
\def\E00{E_{\rm p,0}}
\def\N0{N_{\rm 0}}
\def\F0{F_{\rm 0}}

\def\tt{\tilde{t}}

\def\Dl{d_{\rm  L}}

\def\td{\tau_{\rm dyn}}
\def\ta{\tau_{\rm ang}}

\def\gint{\eta_{\rm int}}
\def\R{{\cal{R}}}
\def\Fb{F_{\rm bol}}

\newcommand{\ltsima} {$\; \buildrel < \over \sim \;$}
\newcommand{\gtsima} {$\; \buildrel > \over \sim \;$}
\newcommand{\lta} {\lower.5ex\hbox{\ltsima}}
\newcommand{\gta} {\lower.5ex\hbox{\gtsima}}

\slugcomment{Date: \today}

\begin{document}

\title{GAMMA-RAY BURST SPECTRA AND LIGHT CURVES AS SIGNATURES OF A RELATIVISTICALLY
EXPANDING PLASMA}

\author{Felix Ryde and Vah\'e Petrosian}

\affil{Center for Space Science and Astrophysics, Stanford University,
    Stanford, CA 94305}

\begin{abstract}
Temporal and spectral characteristics of prompt emission of
gamma-ray burst (GRB) pulses are the primary observations for
constraining the energizing and emission mechanisms. In spite of
very complex temporal behavior of the GRBs, several patterns have
been discovered in how some spectral characteristics change during
the decaying phase of individual, well defined long ($>$ few
seconds) pulses. In this paper we compare these observed
signatures with those expected from a relativistically expanding,
shock heated, and radiation emitting plasma shell. Within the
internal shock model and assuming a short cooling time, we show
that the angular dependence in arrival time from a spherical
expanding shell can explain the general characteristics of some
well defined long GRB pulses. This includes the pulse shape, with
a fast rise and a slower decay, $\propto (1+t/\tau)^2$, where
$\tau$ is a time constant, and the spectral evolution, which can
be described by the hardness-intensity correlation (HIC), with the
intensity being proportional to the square of the hardness
measured by the value of the peak, e.g. $\Ep$ of the $\nu F_\nu$
spectrum. A variation of the relevant time scales involved (the
angular spreading and the dynamic) can explain the broad, observed
dispersion of the HIC index. Reasonable estimates of physical
parameters lead to situations where the HIC relation deviates from
a pure power law; features that are indeed present in the
observations. Depending on the relative values of the rise and
decay times of the intrinsic light curve, the spectral/temporal
behavior, as seen by an observer, will produce the hard-to-soft
evolution and the so called tracking pulses. In our model the
observed spectrum is a superposition of many intrinsic spectra
arriving from different parts of the fireball shell with varying
spectral shifts. Therefore, it will be broader than the emitted
spectrum and its spectral parameters could have complex relations
with the intrinsic ones. Furthermore, we show that the softening
of the low-energy power-law index, that has been observed in some
pulses, can be explained by geometric effects and does not need to
be an intrinsic behavior.

\end{abstract}

\keywords{gamma rays: bursts---gamma rays: theory---relativity}

\section{INTRODUCTION}

The mechanism underlying the prompt $\gamma$-radiation in
gamma-ray bursts (GRBs) is still an unsolved puzzle.  There is,
however, a growing consensus about some aspects of it.  The large
energies and the short time scales involved require the
$\gamma$-rays to be produced in a highly relativistic outflow, an
expanding fireball.  In the standard fireball model $\gamma$-rays
arise from shocks internal to the outflow at a distance of $R \sim
10^{13}-10^{17}$ cm from the initial source. The episodic nature
of the outflow causes inhomogeneities in the wind (or shells) to
collide and thus creating the shocks.  These tap the bulk kinetic
energy and transform it into random energy of leptons which
radiate.  The dominant emission mechanisms are most probably
non-thermal synchrotron \citep{tavani, LP01} and/or inverse
Compton emission \citep{pan00}, but there have been other
suggestions, for instance, thermal, saturated Comptonization
\citep{liang}.

The fundamental process of a burst is thus an individual shock
episode which gives rise to a pulse in the $\gamma$-ray light
curve.  Superposition of many such pulses create the observed
diversity and complexity of light curves \citep{fish94}. The
spectral and temporal characteristics of these pulses hold the key
to the understanding of the prompt radiation of GRBs. However,
there is no consensus on what effects lie behind the observed {\it
pulse} shapes and their temporal and spectral evolution.

The overall spectra of most GRBs can be described by a simple
broken power law with a low and a high energy index, say $\alpha$
and $\beta$, and a break energy $E_b$. Often $\alpha \geq -2$ and
$\beta \leq -2$, so that the $\nu F_{\nu}$ or $EF_E$ spectrum
peaks at a photon energy $E_p\sim E_b$.  The total light curves of
GRBs, on the other hand, are very diverse and not readily
describable by a simple formula. Nevertheless, many
attempts have been made to decompose the complex light curves into
pulses and
analyze their characteristics \citep{norris, lee1, lee2}. No
simple patterns have emerged from these studies of the population
as a whole.  However, some relations have emerged from
investigations of GRBs with simple light curves; those described
by a single pulse or a few, well separated pulses \citep{kar95,
RS00, RS02, BR01} (hereafter BR01). The pulse shapes and evolution
of spectra seem to obey some simple relations. Motivated by these
results, in this paper we explore possible explanations for these
behaviors.

Several different possibilities exist.  The simplest scenario is
to assume an impulsive heating of the leptons and a subsequent
cooling and emission.  The rise phase of the pulse is attributed
to the energizing of the shell which we will refer to as the {\it
dynamic time} and the decay phase reflects the cooling and its
time scale. The instantaneous spectrum reflects the cooling of the
lepton distribution.  The primary problem with this interpretation
is that, in general, the cooling time for the relevant parameters
is too short to explain the pulse durations and the resulting
cooling spectra are in drastic disagreement with the above
observed form \citep{ghis}. A more plausible model is one where
the pulse duration is set by the dynamic time of say the shell
crossing, which could be much larger than the microscopic
acceleration and/or emission-cooling times. In this case there is
a continuous acceleration of particles during shell crossings; the
acceleration and the cooling occur {\it in situ} and
simultaneously and give rise to the observed behavior.  The pulse
shape then is a reflection of the energizing mechanism of the
electrons.  A third possibility is that the above picture operates
only during the rise phase of the pulse and that the decay shape
is due to geometric and relativistic effects in an outflow with a
Lorentz factor of $\Gamma {\gta} 100$.  The curvature of the
fireball shell will make radiation, emitted off the line of sight
(LOS, for short) delayed and affected by a varying relativistic
Doppler boost,  due to the different light paths the photons have
to travel.

The aim of this paper is to investigate to what extent, and how,
the last model affects the observed light curve and spectral
evolution during the individual pulses; in particular to determine
whether the resultant behavior can explain the observed relations
found for simple pulses mentioned above. We want to point out that
the discussion of the observable signatures due to the curvature
effect is independent of the process underlying the intrinsic
pulses of radiation. Colliding shells and internal shocks are one example.
However, other possibilities exist, for instance, as \citet{lyu} pointed out,
if the outflow is Poynting-flux dominated, the intrinsic
radiation could be caused by a non-linear breakdown of large-amplitude
electromagnetic waves at a distance of approximately $10^{14}$ cm
from the progenitor. Furthermore, we emphasize that
the description is for individual emission episodes, i.e. single
pulses and one must bear in mind the possibility that these
actually could consist of several heavily overlapping pulses; see further
discussion in BR01 and \citet{norris}. The observations relevant to our
discussion will be described in \S \ref{sec:obsstat} and the
appropriate time scales in \S \ref{sec:cond}. In \S
\ref{sec:radiation} we derive the spectral and temporal structure
expected in a simplified version of the proposed model. A more
realistic model including both the dynamic and curvature effects
is discussed in \S \ref{sec:broad}. Some other complications and
caveats are discussed in \S \ref{sec:caveats} and a brief summary
and discussion of the conclusions are given in \S \ref{sec:disc}.

In the following, primed quantities are evaluated in the comoving
frame at rest with the outflowing material in the shock front. The
rest frame will denote the inertial frame at rest with the
progenitor.  The cosmological time dilation and spectral redshift,
which are constant factors for individual pulses and bursts, will
be ignored.

\section{OBSERVATIONAL SIGNATURES}
\label{sec:obsstat}

In this section we describe the observational signatures of GRBs
with simple light curves that we wish to explain. Most GRBs
exhibit complex light curves and only a small fraction have
sufficiently long (say longer than few seconds) and smooth pulses
to allow for detailed temporal and spectral investigations. Many
studies have analyzed such small samples of bursts and pulses and
drawn some important conclusions, which give us clues to the
underlying physical processes for the creation of these pulses.

\subsection{Hardness Intensity Correlation (HIC)}

\begin{figure}[]
\epsscale{0.6}
 \plotone{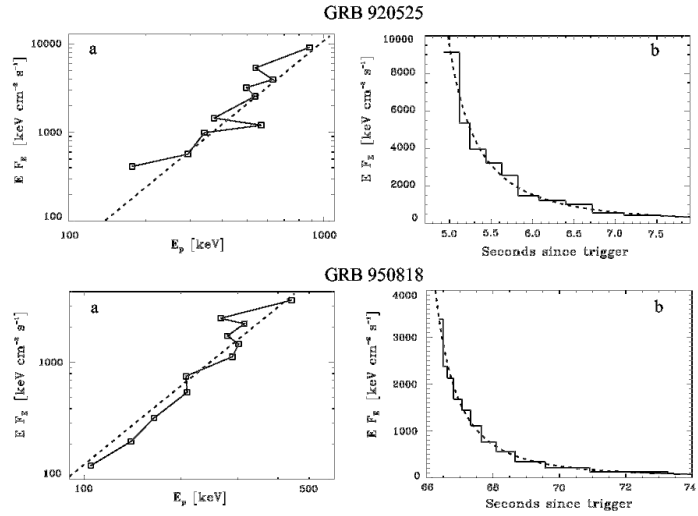}
 \figcaption{Examples of GRBs (squares and histograms) which  follow
the relations in equation (1) and (2) (dashed lines). (a) the HIC
relations and (b) the pulse shapes for the decay phases of
GRB~920525 (trigger 1625) with $\eta=2.4 \pm 0.5$ and $d= 2.0 \pm
0.4$ and of GRB~950818 (trigger 3765) with $\eta=2.2 \pm 0.3$ and
$d= 2.1 \pm 0.4$.
 \label{fig:2cases}}
\end{figure}

\begin{figure}[]
\epsscale{0.6}
 \plotone{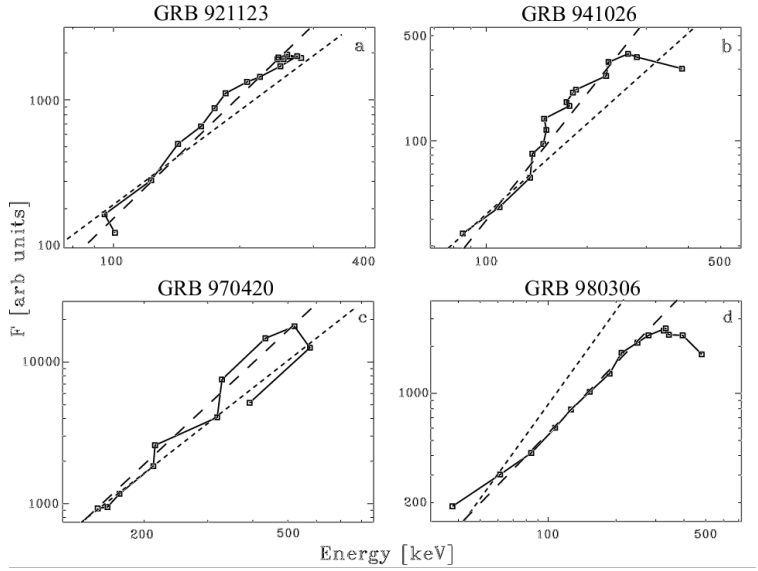}
 \figcaption{Hardness-intensity correlations (HICs) for four BATSE triggers
with $\eta$ indexes deviating significantly from $\eta=2.0$.
 (a) GRB 921123 (trigger 2067) $\eta = 2.60 \pm 0.15$
 (b) GRB 941026 (trigger 3257) $\eta = 2.80 \pm 0.20$
 (c) GRB 970420 (trigger 6198) $\eta = 2.38 \pm 0.15$
 (d) GRB 980306 (trigger 6630) $\eta = 1.49 \pm 0.04$.
 (see BR01 for details). The long-dashed lines show these power-law fits
 to the pulse decay-phase HICs.  Note that the rise-phase HIC is included
 in the figures. The short-dashed lines show the $\eta = 2$ power-law
 showing that the later phases of the pulse in (a,b,c) obey
 this relation (see discussion in \S 5).
 \label{fig:moreHICs}}
\end{figure}

\begin{figure}[]
\epsscale{0.5}
 \plotone{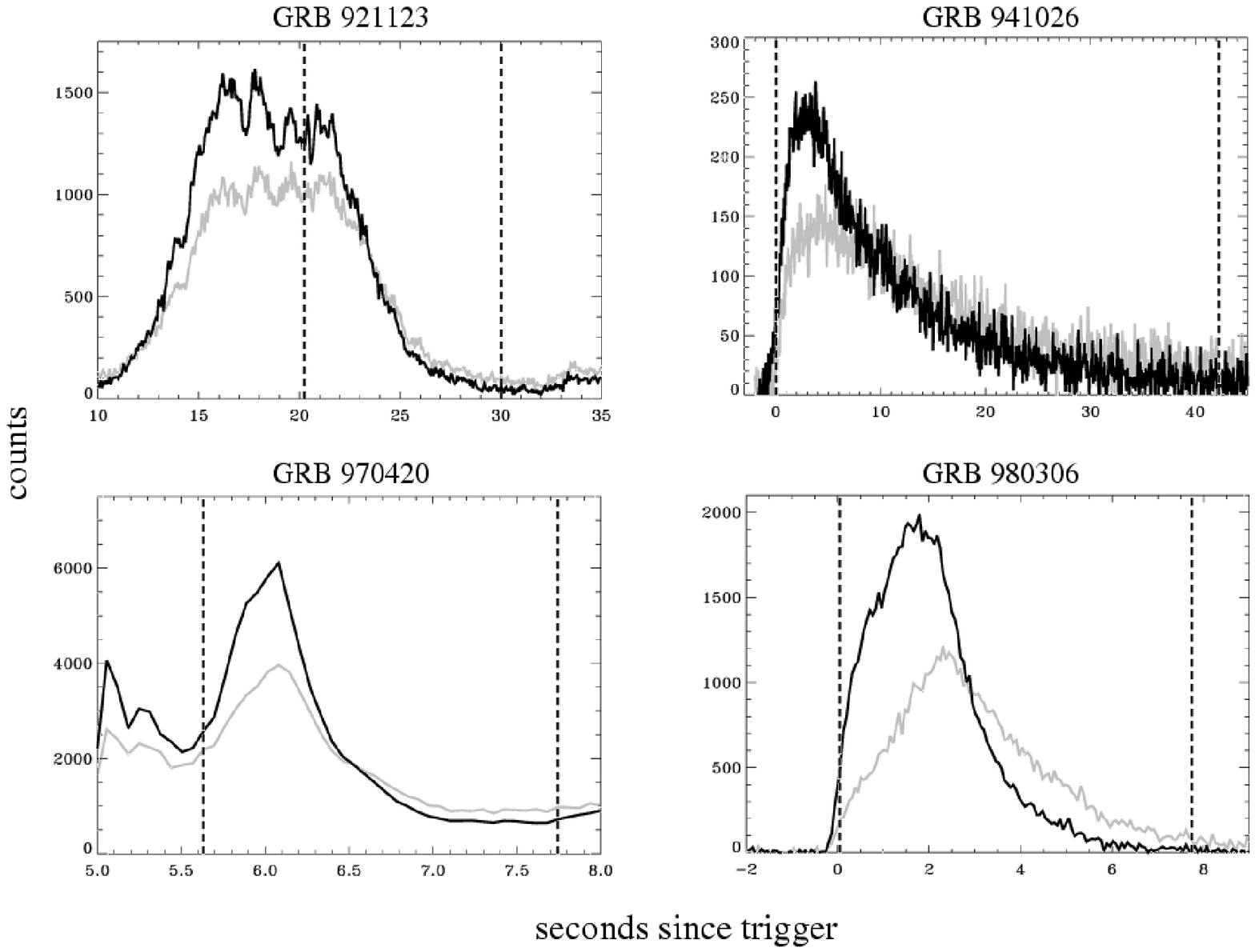}
 \figcaption{Background-subtracted, count light-curves (64 ms resolution)
 for the four triggers in Figure \ref{fig:moreHICs}.
 For each trigger the light curve for two different spectral bands are shown;
 in gray, BATSE channel 2 (55 - 110 keV) and, in black, BATSE channel 3
 (110 - 325 keV). The time interval used for the HICs in Figure
 \ref{fig:moreHICs} are indicated by the dashed lines.
 \label{fig:moreLCs}}
\end{figure}

One of the most important observational signatures is the
correlation between the bolometric energy flux, $\Fb(t) \equiv
\int F(E,t) \, dE$  and the hardness of the spectrum during the
time evolution of individual pulses, where, $F(E,t)$ is the flux
of the energy (not photon) spectrum. This relation, referred to as
the hardness-intensity correlation (or {\it HIC} for short), was
first discovered by \citet{gol83}, using the temperature from
thermal spectral fits as the measure of the hardness. Subsequent
observations have shown that the thermal spectra do not provide a
good fit to a majority of GRBs. More recently BR01, representing
the 'hardness' of the spectrum by the peak photon energy $\Ep$ of
the $\nu F_{\nu} \equiv E F_{\rm E} $ spectrum, find the simple
power-law relation

\begin{equation}
 \Fb=F_{\rm bol, 0} (E_{\rm p}/\E00)^{\eta}, \label{HICobs}
\end{equation}
where $\E00$ and $F_{\rm bol, 0}$ are  some fudicial values of the
peak photon energy and the bolometric energy flux, usually taken
at the beginning of the decay phase.  The time evolution of the
spectrum is mostly from hard and intense to soft and dim.  Figures
\ref{fig:2cases}a and \ref{fig:moreHICs} show examples of the HIC
for pulse decays observed by the Burst and Transient Source
Experiment (BATSE) on the Compton Gamma-ray Observatory ({\it
CGRO}).  More examples can be found by BR01. The distribution of
the power-law index $\eta$ of long GRB pulses is somewhat broad.
The original study by \citet{gol83}, using a thermal fit and $kT$
for $\Ep$, found the power law index to vary between $1.5-1.7$.
Kargatis et al. (1995) studied 26 GRBs with prominent pulses and
found a power-law HIC behavior for 28 pulse decays in 15 of these
bursts. Their distribution was centered on $\eta=1.7$.  BR01
studied a sample of 82 GRB pulse decays and found them to be
consistent with a power-law HIC in, at least, 57\% of the cases
and for these found $\eta = 2.0 \pm 0.7$ (see their Figure 3,
depicting the sample of 47 pulses with good power-law HICs.) An
example of a pulse in this sample with  $\eta < 2$ is shown in
Figure \ref{fig:moreHICs}d. Other such examples can be found in
Figures 4c, 6c,f in BR01. In the BR01-sample, 11 cases have $\eta
>  2.0$ by two standard deviations or more. All of these differ
somewhat from a perfect power-law with most having a concave shape. Three
examples of these are given in Figure \ref{fig:moreHICs}a, b, and
c. Here, whenever possible, the HIC relation for the rise phase is included
(see Fig. \ref{fig:moreLCs} for the time interval used in the
different cases). In Figures \ref{fig:moreHICs}b and \ref{fig:moreHICs}d we
observe a monotonic evolution of $\Ep$ for both the rise and the
decay phase, which is characteristic for the so-called {\it
hard-to-soft} pulses, while  Figure \ref{fig:moreHICs}c illustrates
the so-called {\it tracking} pulses where $\Ep$ and $\Fb$
track each other over the whole pulse, albeit with a slight shift
in time \citep{ford95}.

One essential difference in the BR01 study compared to the earlier
ones is the use of a {\it bolometric} flux measure (see also
\citet{RBS00}) in which the $\nu F_{\nu}$ value at the peak is
used as a measure for this flux. This method was shown to be
better than integrating the spectral flux over the BATSE band, as
long as the peak of the $\nu F_\nu$  spectrum is in the BATSE
window and the power-law indexes, $\alpha$ and $\beta$, do not vary
significantly throughout the pulse. BR01 also showed that the
average distribution of the HIC index is the same whether the flux
was integrated over the observed band or estimated by the peak of
the $\nu F_{\nu}$ flux, indicating that the bolometric correction
for the {\it energy} flux in most cases does not alter the
outcome. The necessity of including a bolometric correction is
more important for the photon number flux than for the energy flux
and is especially true for spectra with soft low-energy photon
distributions. The usefulness of this method is further developed
in \citet{BRVS}. The observational results and data used in this
paper rely on the BR01 approach, so that the discussion will be
instrument-independent as long as one can find a bolometric
correction for the data from the instrument used.

\clearpage
\clearpage

\clearpage

\subsection{Pulse Decay Shape}

Investigations of the light curve (over the whole, observed spectral
range) during pulse decay phases by
\citet{RS00} have shown that the observed photon flux, $F(E)/E$ or
$F(\nu)/\nu$, can often be described by a reciprocal function in
time.  \citet{RKL} derive the  corresponding decay shape for the
bolometric energy flux ($\Fb$), which in general can be described
by
\begin{equation}
\Fb(t) = F_{\rm bol, 0} \left(1+\frac{t}{\tau} \right)^{-d}.
\label{ekv:flux}
\end{equation}
There is, however, an ambiguity in fitting such a light curve to
the data. The values of the power-law index $d$ and the time
constant $\tau$ are coupled  and, consequently are often not well
constrained by the fitting. Furthermore, a subjective judgement
must be made in choosing the transition moment between the rise
and the decay phases. As a result the pulse shape fitting does not
provide as clear a signal as the HIC relation. This ambiguity is
illustrated by F. Ryde, D. Kocevski, et al. (in
preparation), who revisit the BR01 sample and analyze the
deconvolved light curves and fit the light curves with equation
(\ref{ekv:flux}). The parameters are indeed unconstrained in
approximately half of the cases. Two of the constrained cases  are
shown in Figure \ref{fig:2cases}b. \citet{KL} and \citet{RKL}
introduce an approach to overcome this ambiguity by defining
various analytical shapes for the whole pulse that include the
rise phase (whenever present) and asymptotically approaches
equation~(\ref{ekv:flux}) in the decay phase. \citet{KL} studied
a sample of 22 pulses with good rise phases and found $d= 2.2 \pm 0.7$.

A corollary of the above two relations is that $\Ep$ also decays
following the above form with a different exponent.

\begin{equation}
\Ep (t) = \Epkmax \left(1+\frac{t}{\tau} \right)^{-d/\eta}.
\label{Epevol}
\end{equation}

\subsection{Change of Pulse Shape with Spectral Band}

In this paper we will mainly focus on the observed behavior
of the spectral evolution as described by the HIC and the
shape of the pulse decay phase. There are, however, several
other approaches that can be used to present the observed evolution. One
way is to study the change in pulse shape and width over
different spectral bands and the time lag between these bands
(see, e.g., \citet{norris2}). To illustrate this we present
in Figure \ref{fig:moreLCs}  the light curves of
the four cases shown in Figure \ref{fig:moreHICs}. For each
case we show the light curve in two energy bands. For more detailed
discussion on this approach we refer to F.Ryde, D.Kocevski, et al.
(in preparation).

\section{TIME SCALES}
\label{sec:cond}

There are many time scales in a GRB.  In this paper we are
interested in the effects of the curvature of the emission front
and the time scale it introduces in the pulse shape.  In order to
find the importance of this effect its time scale should be
compared to the radiative cooling time and the dynamic time for
the crossing (or merger) of two shells.

Estimation of {\it the cooling time}, which we assume is
comparable or longer than the {\it acceleration time}, requires a
knowledge of the emission and energy loss processes of electrons.
The details of the acceleration mechanism are not known, but it is
generally assumed that the dissipation of the kinetic energy of
the fireball gives rise to a power law distribution of electrons.
The minimum random Lorentz factor of the electrons is proportional
to the relative Lorentz factor; $\gamma_{\rm{e}} = \xi_{\rm{e}}
(m_{\rm{p}}/m_{\rm{e}})(\Gamma_{\rm{rel}}-1)$, where $m_{\rm{e}}$
and $m_{\rm{p}}$ are the electron and the proton masses,
respectively, and $\Gamma_{\rm{rel}} = \Gamma_{\rm{1}}
\Gamma_{\rm{2}} (1-\beta_1 \beta_2)  \sim \Gamma_{\rm{1}}/(2
\Gamma_{\rm{2}})$, which is of the order of $10$ or lower, is the
relative Lorentz factor between the two interacting shells
($\Gamma_{\rm{1}} > \Gamma_{\rm{2}}$). $\xi_{\rm{e}}$, and
$\xi_{\rm {B}}$ introduced below, are the fractions of the
post-shock, random energy density that resides in the electrons
and the magnetic field.  The magnetic field strength is given by
$B = (32 \pi \xi_{\rm{B}} n' m_{\rm{p}} c^2
\Gamma_{\rm{rel}})^{1/2} \sim 2 \times 10^4$ Gauss, where we
assume equipartition and a comoving electron density $n' =
L_{\rm{kin}}/ 4 \pi R^2 \Gamma^2 m_{\rm{ p}} c^3 \sim 2 \times
10^{10}$ cm$^{-3}$, for an assumed kinetic luminosity of
$L_{\rm{kin}} \sim 10^{53}$ erg/s, distance $R$ of $10^{15}$ cm,
and a shock compression ratio of 4.

In the rest frame of the progenitor the cooling time scale for a
particle of comoving energy $E'$ and loss rate $\dot{E'}$ is
dilated to $\tau_{\rm loss}= \Gamma E'/\dot{E'} $. As the emitting
material is moving towards the observer the rest-frame time scales
will be compressed by a factor $(1-\beta)^{-1} \sim  2 \Gamma^2$
in the observer frame (see below). If the pulse shape is
determined solely by the radiative cooling process then, for the
above parameters and assumptions, the observed decay time scale of
the pulse will be

\begin{eqnarray}
\tau_{\rm cool}&=&\frac{E}{2\Gamma |\dot{E_{\rm s}}+
{\dot{E_{\rm{IC}}}}|}=\frac{3 \pi m_{\rm{e}}c} {\Gamma \sigma_{\rm
{T}} \gamma_{\rm{ e}}
B^2(1+U_{\rm{rad}}/U_{\rm{B}})}\nonumber\\
&\sim&6 \times 10^{-5} \, {\rm s} (\Gamma_{\rm{rel}}-1)^{-2}
\left(\frac{R}{10^{15} {\rm cm}}\right)^{2}
\left(\frac{\Gamma}{100} \right)
 \left(\frac{L_{\rm kin}}{10^{53}} \right)^{-1}
\label{tcool}
\end{eqnarray}
Here subscripts s and IC refer to the cooling rates due to
synchrotron and inverse Compton losses in radiation and magnetic
field energy densities of $U_{\rm rad}$ and $U_B=B^2/(8\pi)$. This
is a much shorter time scale compared to the dynamic and curvature
times discussed below.

\begin{figure}[]
\epsscale{0.5}
 \plotone{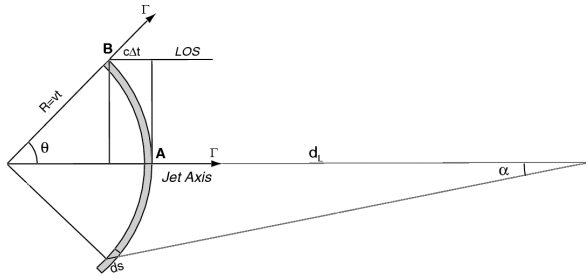}
 \figcaption{
Geometrical sketch of the visible part of the fireball. Due to the
relativistic aberration of light, the main part of the radiation
comes out within a small cone with half opening angle $\theta \sim
\Gamma^{-1}$. Note that the line of sight (or LOS) and the jet
axis do not need to be parallel. The photons from A are boosted by
$2 \Gamma$ but those from B, which are delayed by $\Delta t = R/c
(1- \mu)$, are boosted by $[\Gamma (1- \beta \mu)]^{-1}$, where
$\mu= \cos \theta$. \label{figgeom}}
 \end{figure}

{\it The curvature time scale} arises from relativistic effects in
a sphere expanding with a high bulk Lorentz factor $\Gamma$.  Due
to the curvature of the shell there will be a time delay between
the photons emitted simultaneously in the comoving frame from
different points on the surface.  Figure \ref{figgeom} shows the
geometry of the situation.  Due to the relativistic aberration of
light, isotropically emitted radiation in the comoving frame will
be beamed into a cone with opening angle $\theta \sim
\Gamma^{-1}$.  Only photons emitted from the fireball surface
within a narrow  cone of opening angle $\sim \Gamma^{-1}$ around
the LOS will be detectable by the observer.  The typical time
delay is thus $\ta = R[1-\cos(1/ \Gamma)]/c$ which for large
$\Gamma$ is approximately $R/(2\Gamma^2) c$.  This gives a lower
bound for the observed duration of a pulse:

\begin{equation}
\tau_{\rm ang} = 1.7 {\rm   s}\left(\frac{R}{10^{15} {\rm
cm}}\right) \left(\frac{\Gamma}{10^2}\right)^{-2}.
\label{ekv:tang}
\end{equation}

{\it The dynamic time scale} for a single pulse is the actual
crossing (or merger) time of one shell with another.  Often the
shell collision is assumed to be an inelastic collision and the
merged shells expand as a single shell.  The shell crossing time
is $\td' = \Delta'/v'_{\rm sh}$, where ${v'_{\rm sh}}$ is the
velocity of the shock in the comoving frame of the preshocked
flow. The initial value, and the evolution with radius, of the
shell width $\Delta '$ are not well understood and depend, among
other factors, on the structure and internal dynamics of the shell
and on its interaction with the external medium.  If the Lorentz
factor is constant over the shell, or if the shell is confined by
some mechanism, then the shell width will be independent of radius
and time, $\Delta=\Delta_0 $. But if there exists a differential
flow with a faster leading and a slower trailing edge with
velocities $\beta_{\rm r}$ and $\beta_{\rm s}$, and corresponding
Lorentz factors $\Gamma_{\rm r}$ and $\Gamma_{\rm s}$,
respectively, then

\begin{equation} \Delta(R)= R(\beta_{\rm r}-\beta_{\rm s})=
\frac{R}{2\Gamma_{\rm s}^2} \left[1-\left(\frac{\Gamma_{\rm
s}}{\Gamma_{\rm r}}\right)^2 \right]\sim \frac{R}{2\Gamma_{\rm
s}^2},\label{equ:delta}
\end{equation}
 where for the last relation we have assumed that $\Gamma_{\rm s} \ll \Gamma_{\rm
r}$. In the comoving frame  $\Delta'(R)= R/(2\Gamma_{\rm s})$.
This relation will be applicable at radii $R>R_{\rm b}= \Delta_0
\Gamma^2$,  where the spread will exceed the initial width.  The
shell crossing time is then $\tau_{\rm dyn}'= \Delta'/v'_{\rm
sh}=R/(\Gamma_{\rm s} v'_{\rm sh})$.  In the observer frame these
times will be time-dilated and affected by the motion of the
fireball towards the observer:
\begin{equation}
\tau_{\rm dyn}= \frac{\tau_{\rm dyn}'}{2 \Gamma} =
\frac{R}{4\Gamma_{\rm s}\Gamma v_{\rm sh}'}= 1 \, {\rm s}\,
\beta'^{-1} _{\rm sh} \left(\frac{R}{10^{15}}\right)
\left(\frac{\Gamma }{100 } \right)^{-2} \left(\frac{\Gamma
}{\Gamma_{\rm s}} \right). \label{tcross}
\end{equation}

\noindent
 For the assumptions described above the dynamic
time scale is somewhat shorter than (and has a similar dependence on
$R$ and $\Gamma$) as the curvature time scale; $\tau_{\rm ang}\sim
2(\Gamma_{\rm s}/\Gamma)\tau_{\rm dyn}$. However, at smaller
values of $R<R_b$, where the spreading is negligible, the dynamic
time scale may exceed the angular one by $R_b/R$.  Since, in
general, the pulse width is proportional to $R$, the latter
situation is more likely to arise in shorter pulses.

\begin{figure}[]
\epsscale{0.5}
 \plotone{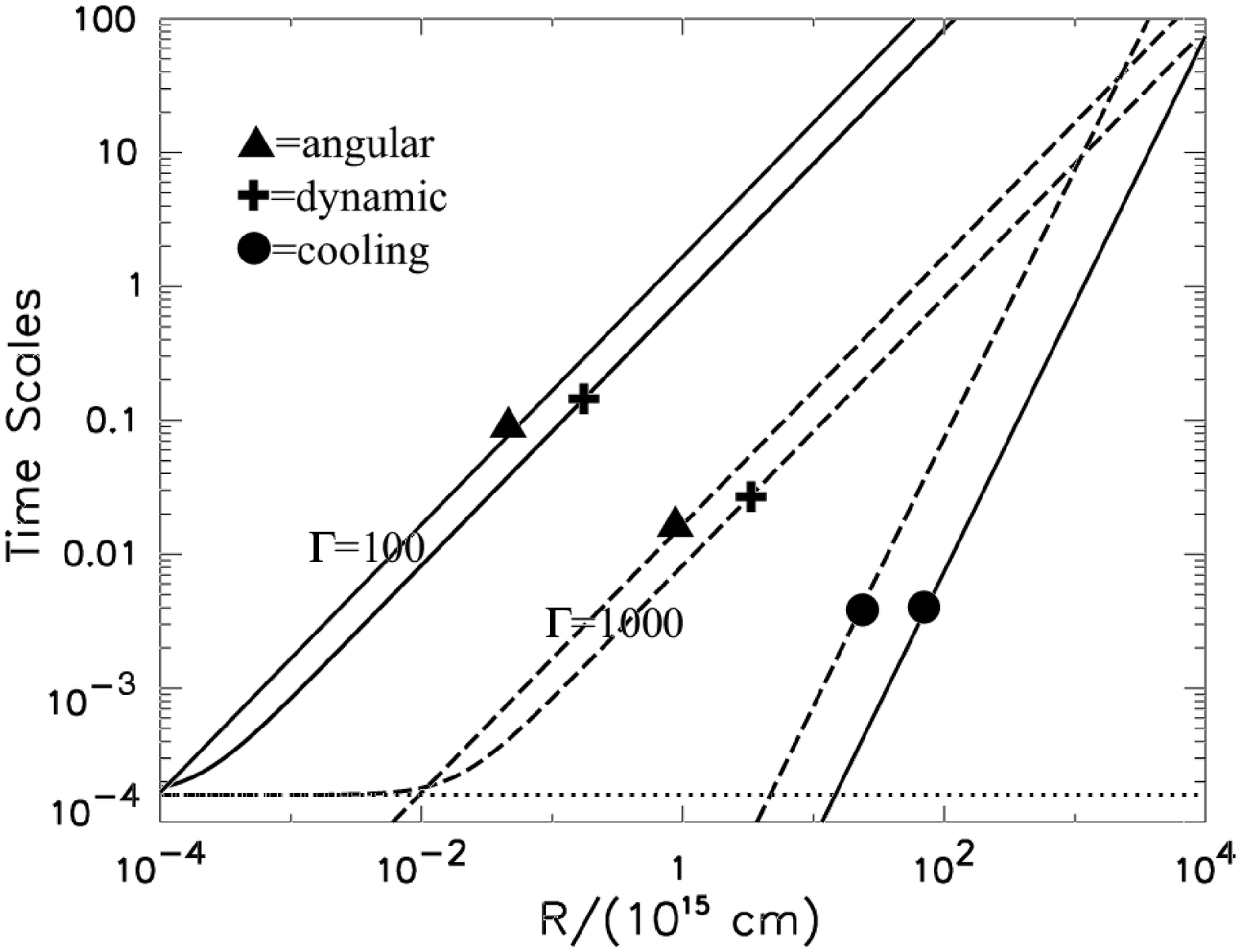}
 \figcaption{
The radial dependence of the various time scales. The solid lines
are for a bulk Lorentz factor of $\Gamma= 100$, while the dashed
lines are for $\Gamma= 1000$. For the cooling time we have used
eq. (\ref{tcool}) with $\Gamma_{\rm rel}=10$ and a kinetic
luminosity $L_{\rm kin}=10^{53}$ erg/s, and the angular time is
given by equation (\ref{ekv:tang}). For the dynamic time we have
used equation (\ref{tcross}) with  $\Gamma _s \sim \Gamma = 100$
and $\beta'_{\rm sh} = 1$. If the shell width is constant,
$\Delta=\Delta_0$, then there will be a minimum dynamic time scale
at all $R$ and be represented by a horizontal line in the figure;
Assuming $\Delta_0= 10^{9}$ cm,
 $\td= 1.6 \times 10^{-4}$ s ${\beta'_{\rm sh}}^{-1}(\Gamma/100)^{-1}
 (\Delta_0/10^{9}$ cm).
 \label{figtime scales}}
\end{figure}

Figure \ref{figtime scales} shows the radial dependence of the
three time scales, $\tau_{\rm cool}$(eq.  \ref{tcool}), $\tau_{\rm
ang}$(eq. \ref{ekv:tang}), $\tau_{\rm dyn}$ (eq. \ref{tcross}),
for two values of the bulk Lorentz factor $\Gamma$, for $\Gamma
\sim \Gamma_{\rm s}$, and for $R_b=10^{11}$ cm.  From this we
conclude that in situations with sufficiently narrow shells, the
curvature effects can indeed be the determining factor for the
pulse decay time scale. Correspondingly, if these effects are shown
to be important in the observed pulses, this could set a
constraint on the thickness (value of $\Gamma_{\rm s}/\Gamma_{\rm
r}$) and the spreading of the shell (values of $R_b$ and
$\Gamma_{\rm s}$) and/or on the distance $R$ where the radiation
is emitted. The curvature effect becomes important if $\tau_{\rm
ang} > \tau_{\rm dyn}$. This happens for shell widths $\Delta'
\lta\beta'_{\rm sh} R/\Gamma$. Even in the case with linearly
increasing shell-width the angular effect will be noticeable and
can be totally dominating if the shell is narrower (e.g. if the
shell-width stays constant). \citet{SPM00} simulated internal
shocks and found that with a linear shell broadening, as above,
the angular spreading and the shock crossing times are comparable.

In the following we consider a scenario where the rise phase of a
pulse corresponds to the merging phase or the dynamic evolution,
while the decay phase arises as a purely kinematical effect due to
the curvature of the relativistically expanding shell.  The acceleration
and radiative cooling times are assumed to be much shorter than or
at most comparable to the dynamic time.

\section{RELATIVISTIC KINEMATICS}
\label{sec:radiation}

The radiation from a relativistically expanding plasma sphere
(fireball) will have a unique signature in the observer frame.  We
start by studying the radiation observed from an infinitely thin
spherical shell and thereafter generalize the problem to broader
shells. We assume that the visible part of the fireball is
spherically symmetric and homogeneous, {\it i.e.}, only one
parameter is enough to characterize the properties of a patch of
emission. This type of problem has previously been discussed by, among others,
\citet{Fen}, \citet{Gra99} and  \citet{eri00}.

\subsection{Lorentz Boosting and Spectrum Evolution}

\label{bsec:boost}

The Lorentz boosting factor for transformation from the comoving
frame to the observer frame of photons emitted into the LOS from
different locations on the surface, defined by the angle $\theta
\equiv \arccos \mu$ shown in Figure \ref{figgeom} is
\begin{equation}
{\cal D} (\Gamma,\mu)=\frac{1}{\Gamma (1-\beta \mu)} =\Gamma
(1+\beta \mu') .
\label{boost}
\end{equation}
For small angles this reduces to ${\cal D} (\Gamma)= 2\Gamma$,
which is the boost factor used in discussing GRB emission, when
the angular dependence is not considered important (e.g.  flat
shell perpendicular to LOS).

If we set $\theta=0$ at the point where the flow velocity is
parallel to LOS (point A in Fig.  \ref{figgeom}), then the
difference in light travel time $\Delta t$ between photons emitted
along the LOS from this point and a point at an angle $\theta$
(point B in Fig.  \ref{figgeom}) is $\Delta t = R_0(1-\mu)/c$ ,
which gives $\mu= 1-c\Delta t/R_0$.  Inserting this into equation
(\ref{boost}) we find

\begin{equation}
{\cal D} (\Gamma,\Delta t)=\frac{1}{\Gamma (1-\beta + \beta c
\Delta t/R_0)}.
\end{equation}
For highly relativistic outflows $\beta = 1-(2\Gamma^2)^{-1}$, and

\begin{equation}
{\cal D} (\Gamma,\Delta t)= 2 \Gamma (1+\Delta t/\tau_{\rm
ang})^{-1}, \,\, \, \tau_{\rm ang}\equiv R_0/(2\Gamma^2 c).
\label{Dt}
\end{equation}

An obvious outcome of this is that if the emitted spectra from
different parts of the shell are identical, then the observed
spectrum will be boosted (gradually redshifted) in time by a
factor ${\cal D}$ as different parts of the surface come into
view. In particular the peak energy  will evolve as

\begin{equation}
\Ep(t)= E'_{\rm p} {\cal D}=\frac{\Epkmax}{(1+t/\tau_{\rm ang})},
\label{Et_rel}
\end{equation}
where $E'_{\rm p}$ is the peak energy in the comoving frame (which
is the same at all angles), $\Epkmax=E'_{\rm p}2\Gamma$, and we
have set $\Delta t = t$, which assumes that $t=0$ for radiation
observed from point A in Figure \ref{figgeom}.

\subsection{Energy Flux and the Light Curve}
\label{sec:LC}

We want to determine the pulse shape, in bolometric energy flux,
assuming it to be entirely caused by the above angular dependence
of the Lorentz boost variation. This means we are considering a
situation where $\tau_{\rm dyn}$  and $\tau_{\rm cool} \ll
\tau_{\rm ang}$.   We, therefore, initially assume a simple
isotropic volume emissivity.

\begin{equation}
j'=\Sigma' \, \delta(t'-t_0) \,
\delta(R'-R_0),
\label{emiss}
\end{equation}
where $\Sigma'$ is the total, rest frame, energy surface-brightness.
More complex temporal and spatial distributions will be discussed
in the next section. [We follow the formalism outlined in
\citet{Gra99} and refer to Figure \ref{figgeom}.] We also
introduce the following variables: the luminosity distance $\Dl$,
the viewing angle $\alpha = \sin \theta R/\Dl =  (1-\mu^2)^{1/2}R/
\Dl$ so that $\alpha d \alpha = (R/\Dl)^2 \mu \, d\mu$. Using the
Lorentz invariance of $j/\nu^3$ \citep{RL} we can write  the
observed intensity (neglecting cosmological redshift factors) as

\begin{equation}
I = \int jds=\int j' {\cal{D}}^3 d R/\mu
\end{equation}
and the observed bolometric energy flux as

\begin{equation}
\Fb=\int \mu I d\Omega =\int \int \int j'{\cal{D}}^3 dR
 \sin \alpha d\alpha d\Phi
\end{equation}
where $ds$ is the element of length through the shell along the
LOS, $\Phi$ is the azimuthal angle and $d \Omega$ is the solid
angle of the source as seen by the observer. Assuming azimuthal
symmetry around the jet axis ($\int d \Phi= 2 \pi$) we get

\begin{equation}
\Fb(t) = \frac{1}{2 \Dl ^2} \int_0 ^\infty  R^2 d R \int _{-1}
^{1} \; \mu \, d \mu \, {\cal{D}}^3 (4\pi j').
\end{equation}

Using the transformations $| d R/d R'| = \Gamma^{-1}$, $|d t_{\rm
obs}/dt'| = {\cal D}^{-1}$, and that $d t = (R/c) \, d \mu$ we get
$\delta(t'-t_0) \delta(R'-R_0) = c \, (R \Gamma {\cal{D}})^{-1} \,
\delta(\mu - \mu_0)  \delta(R-R_0)$. The observed flux then can be
written as
\begin{equation}
\Fb(t)= F_0 \mu {\cal{D}}^2(\mu)/\Gamma^2, \,\,{\rm with}\,\, F_0
\equiv {2\pi \Sigma' R_0 c\Gamma}/{\Dl^2}, \label{ekv:F}
\end{equation}
where the time dependence arises from the relation $\mu = 1 -
ct/R_0$. For a highly relativistic outflow (where $\theta _ {\rm
max} \sim 1/ \Gamma$) the projection factor $\mu=\cos \theta$ in
equation (\ref{ekv:F}) is less than $1-(2 \Gamma^2)^{-1}$ and can
be ignored so that $\Fb(t) =  F_0 {\cal D}^2(\mu)/\Gamma^2$. Note
that the above result is independent of whether the outflow is
completely spherical or confined or collimated into a jet of
opening angle $\theta_{\rm jet}$, as long as $\theta_{\rm jet} >
\Gamma^{-1}$.

Now with the help of equation (9), and identifying the time delay
$\Delta t$ with the time $t$ as above, we get
\begin{equation}
\Fb(t) = F_0/(1+t/\tau_{\rm ang})^2.
 \label{Ft_rel}
\end{equation}
\noindent
 This surprisingly  is very similar to the observed
behavior of the small sample of well-isolated pulses described in
\S 2.

\subsection{Flux-Spectrum Correlation; HIC}
\label{sec:HICmodel}

In the $\delta$-function approximation, and if the bulk Lorentz
factor and if the rest frame  flux and spectrum  are independent
of position along the surface of the shell ({\it i.e.} we have a
homogeneous source), then equations (\ref{Et_rel} and
\ref{Ft_rel}) will describe the spectral and  flux evolutions
adequately. This can be translated into a measurable
hardness-intensity correlation if we represent the hardness by the
value of $\Ep$. The observed HIC (eq. [\ref{HICobs}]) follows
directly: $F\propto \Ep^\eta$, with a power-law index of $\eta=2$,
which is equal to the observed average value found in BR01.

Another way to look at this relation is through the time evolution
of the energy \underline{fluence} defined as  ${\cal {E}} (t)
\equiv \int_0^t F(\tt) d\tt$. Using equation (\ref{Ft_rel}) it can
easily be shown that

\begin{equation}
{\cal {E}} (t) = {F_0
 \tau} \left[1-(1+t/\ta)^{-1}\right],
\end{equation}
which when inserted into equation (\ref{Et_rel}) gives

\begin{equation}
\Ep = \E00 - \frac{{\cal E} \E00}{\F0 \ta} ,\label{19}
\end{equation}
which is the so-called hardness-fluence correlation (HFC). The
inverse of its proportionality constant is usually denoted by
$\P0$ \citep{LK96, RS00}, so in this description $\P0 = \F0
\ta/\E00$.

\section{FINITE DYNAMICAL TIME}

 \label{sec:broad}

The above analysis assumes a very short dynamic, as well as
cooling, time scales relative to the curvature time delay.  This
allows the delta function time profile approximation used in \S
\ref{sec:LC}.  However, as mentioned above and shown in Figure
\ref{figtime scales}, the dynamic time $\tau_{\rm dyn}$, which in
our model defines the rise time of the pulse, could be comparable
to (and in short pulses it may exceed) the curvature time scale.
Therefore we must examine the effects of finite $\tau_{\rm dyn}$.

\subsection{Bolometric Light Curves} \label{sec:LC2}

For a shell shining continuously for a finite period of time the
resulting pulse profile will be a convolution of the comoving
frame emissivity time profile, say $j'(t') = \Sigma' f(t')$,  with
$\int f(t') dt' = 1$, and the profile due to the curvature effect
(eq. [\ref{Et_rel}]) obtained from an impulsive intrinsic emission
profile, $f =\delta (t' -t'_0)$.  The emissivity profile, when
transformed to the observer frame, retains its form but its
characteristic time scale is scaled by the boost factor; $\tau_{\rm
dyn}'=\tau_{\rm dyn}{\cal D}$. The resultant pulse profile then
becomes
\begin{equation}
\Fb (t)= F_{\rm 0} \int_t^{+ \infty} f(t - x)dx/(1 + x/\ta)^2
\label{ekv:Fconv}
\end{equation}

\begin{figure}[h]
 \epsscale{0.7}
 \plotone{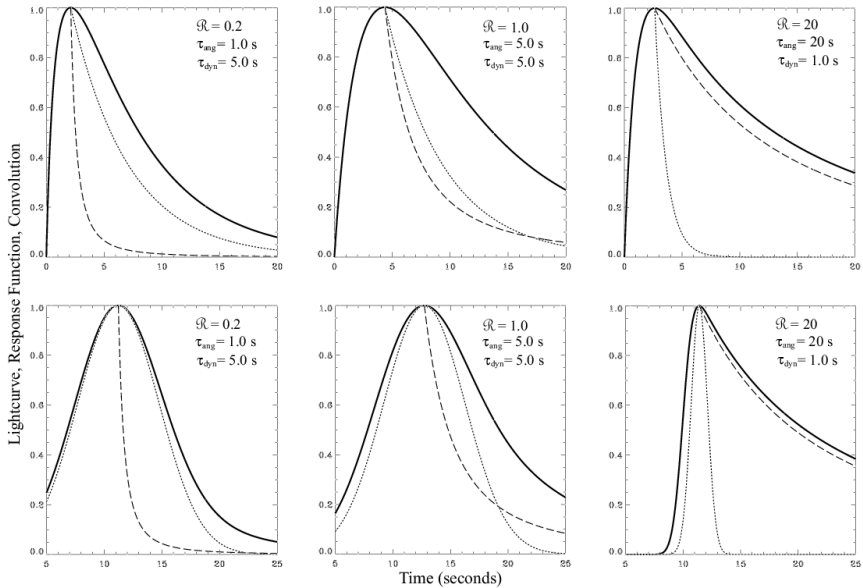}
 \figcaption{Light curve behaviors in the observer frame, with
 varying parameter combinations of the exponential decay
 constant $\td$ and the angular time scale $\ta$. The dotted, dashed, and solid lines
 represent the comoving light curve, the relativistic angular spread function and the
 convolved light curve. Top panels for a one-sided (zero rise time) exponential
 intrinsic light curve  $f \propto \exp (-t/\td), \,\, t>0$.  Note that the
convolved light curve has a finite rise time. The lower panels for
a symmetric Gaussian intrinsic light curve, $f \propto \exp
(-t/\td)^2$, for $- \infty < t < \infty$. The comoving light
curves and the angular spread functions  are normalized to have
the $t=0$ at the peak of the observed (convolved) pulse.
  \label{figLCs}}
 \end{figure}

Figure \ref{figLCs} illustrates the effects of the finite emission
time scale for the intrinsic pulses. In the three top panels the
intrinsic pulses were modeled as decaying exponentials with
time scale $\td$. The shape of the convolved light curve is
determined by the ratio of the intrinsic emission time scale $\ta$
to that of the curvature $\td; {\cal R}=\tau_{\rm ang}/\tau_{\rm
dyn}$.  In general the convolved pulse will resemble a FRED (fast
rise and exponential decay). Different ratios of the time scales
give rise to pulse shapes that encompass the actual observed pulse
shapes by BATSE. For long duration pulses the convolved light
curves will asymptotically reach the $d=2$ form of equation
(\ref{ekv:flux}) and manifest the curvature effect. But for short
and weak bursts this stage may not be observable and one sees a
pulse shape determined by the dynamics of the shell crossing. The
general behavior is well illustrated by these three cases.
However, other intrinsic pulse shapes do give rise to some
variations. For instance, including an intrinsic rise phase will
broaden the rise phase of the convolved light curve, as
illustrated by the bottom panels in which the intrinsic pulses
were modeled by symmetric Gaussians  of half width $\td$.

It should be emphasized again that in  the discussion above, and
in equation (\ref{ekv:Fconv}) in particular, we are dealing with
the {\it bolometric} flux, using the approach introduced in BR01,
who represent this flux by $E F_{\rm E}$ evaluated at $\Ep$. In
general, however, for a finite observation band there will be a
time dependence term beyond what is shown in equation
(\ref{ekv:Fconv}). If, over a finite dynamic time, the spectrum
evolves, {\it i.e.} the $\Ep $ changes, the flux contribution
within the observed band, and consequently  the bolometric
correction will change (the amount of changes
 depending on the values and/or evolution of indexes $\alpha$ and
$\beta$). This will affect the above light curve as well as the
HIC relation described next.  In an approach  different  from
BR01, to account for this issue, \citet{FenRik, Fen} instead
assumed a photon number spectrum described by a single power law
with index equal to the averaged value $(\alpha+\beta)/2$, and
calculated the observed photon flux over a certain band pass which
thus changes in the observer frame with time. The relation between
the photon flux and the peak energy found by this method was studied by
\citet{sod01} who concluded that a long cooling time was necessary
to explain the data.

\subsection{Spectra and HICs}
 \label{sec:cases}

The effects of a finite emission pulse on the spectral behavior are
more complicated. This is because the instantaneous spectra, as
seen by the observer, will be a superposition of spectra from
several annuli emitted at different times having different
$\mu=\cos\theta$, boost factor ${\cal D}(\mu)$, and perhaps
different spectral hardness (say different $\Ep$) and shape
(different indexes $\alpha$ and $\beta$). Let us assume that the
intrinsic spectrum, as observed in the rest frame, can be
described by two spectral indexes $\alpha$ and $\beta$ and a break
or peak energy $\Ep$, say a \citet{band} function, ${\cal
B}_{\alpha, \beta}(E/\Ep)$, with $\int_0^{\infty} {\cal B}\,dE
=1$. To simplify matters, we further assume that $\alpha$ and
$\beta$ are constant so that the only spectral variation is due to
changes in $\Ep$. We will represent the evolution of $\Ep$ by an
intrinsic HIC relation with index $\gint$, $F \propto {E_{\rm
p}}^{\gint}$, which means that $\Ep (t) = E_{\rm p,0}
[f(t)]^{(-1/\gint)}$. The observed flux spectrum at any time can
thus be written as
\begin{equation}
F(E,t) = F_0 \int_t^{\infty} \frac{f(t-x)}{(1+x/\ta)^2} {\cal
B}_{\alpha, \beta}\left(\frac{E}{\Ep(t,x)}\right) \, dx, \,\,\,
 \label{ekv:FconvE}
\end{equation}
with
\begin{equation}
\Ep(t,x)=E_{\rm p,0}(1+x/\ta)^{-1}[f(t-x)]^{(-1/\gint)}
\end{equation}
When integrated over all energy, equation (\ref{ekv:FconvE})
reduces to equation (\ref{ekv:Fconv}). Note that the resultant
spectra or light curves depend primarily on the ratio of the time
scales, ${\cal R}=\ta/\td$ defined above. Secondary factors are
the intrinsic pulse shape and HIC relation.

Before showing the resultant HIC for different cases we first
discuss the details of the formation of the observed spectra. For
the purpose of illustration we assume $\alpha=-2/3, \beta=-2.5$
and three different values for the index $\gint$. To simplify the
description of how the contributing spectra make up the integrated
spectrum we start with the case of an intrinsic light curve with
an abrupt rise phase and a simple exponential decay phase:
$f(t)={\td}^{-1} \exp(-t/\td), \,\, {\rm for} \,\, t>0$. The
effects of intrinsic pulses with a finite rise phase are discussed
below. Note that an intrinsic pulse with only a decay phase will
still produce a rise phase for the observer due to the convolution
described above. For such an intrinsic pulse, the upper limits of
the integrals in eq. (\ref{ekv:Fconv}) and (\ref{ekv:FconvE}) are
zero, so that at any given time $t$ the observer receives signals
from annuli extending from $\theta=0$ to $\theta=\theta_{\rm
max}(t)$ or $1 < \mu < \mu_{\rm min}=(1-\beta c t/R_0)$. Different
angles $\theta$ sample different stages of the intrinsic spectral
evolution due to the angular spreading of the signal, with the
spectrum from $\theta_{\rm max}(t)$ being that emitted at the
beginning of the pulse and the spectrum from $\theta=0$ reflecting
that of a later time in the pulse. Equation (\ref{ekv:FconvE}) gives
the convolved spectrum as a superposition of different spectra
from different angles or different times.

In the absence of the curvature effect ({\it i.e.} for a flat
shell, $R_0 \rightarrow \infty$, or $\ta$ and ${\cal R}
\rightarrow 0$) one would observe the intrinsic HIC relation. It
follows then that for a finite $\ta$ but for ${\cal R} \ll 1$,
this is the HIC relation that will be observed except for a short
time $t \sim \ta$ at the beginning (and the rising phase) of the
observed pulse. In the opposite limit of ${\cal R} \gg 1$, the
delta function approximation result with the HIC relation $\Fb
\propto \Ep^2$ is obtained, again except for a short time $t \sim
\td$ at the beginning (and the rising phase) of the observed
pulse.

\begin{figure}[]
\epsscale{0.5}
 \plotone{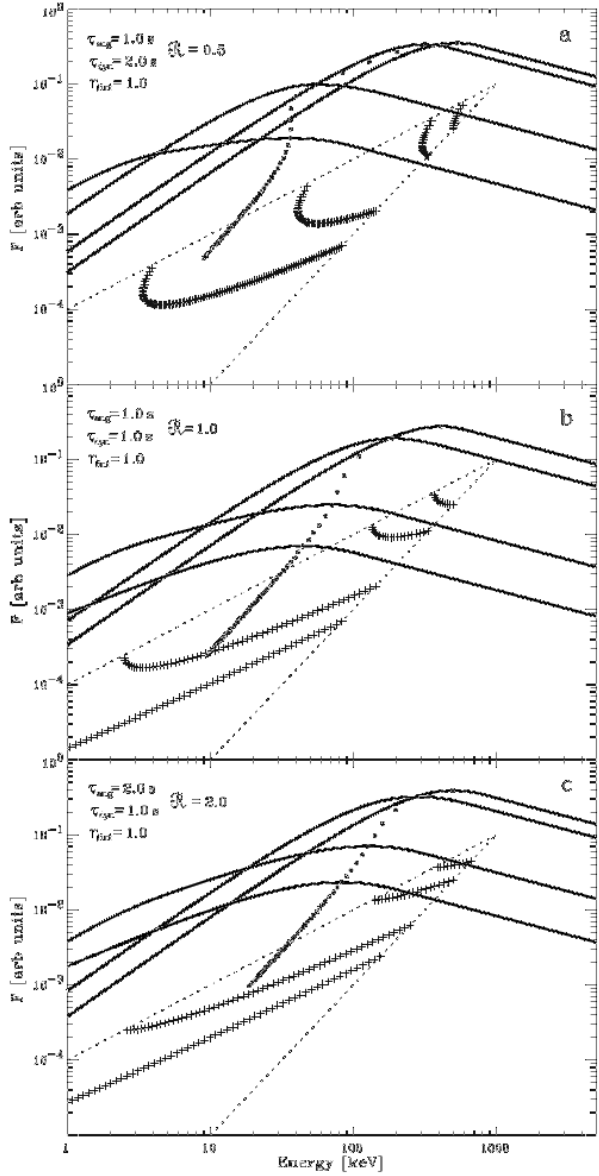}
\caption{ Observer frame spectra (solid lines) at four different
times (1, 2, 6 and 11 s; where the first light is detected at
$t=0$ s). These spectra are found by integrating the intrinsic
spectra over $\theta$. The positions in the $F-\Ep$ plane of the
peaks of some of these contributing spectra are indicated by large
crosses. The upper and lower dashed lines represent, respectively
the locii at $\theta=0$ which follow the intrinsic evolution of
the HIC relation (here assumed to be  a power law with intrinsic
HIC index, $\gint=1.0$), and  at $\theta=\theta_{\rm max}$ obeying
the HIC relation expected from the curvature effect (with a
power-law with index $\eta = 2$). The circles show the HIC
relation of the convolved spectra in time steps of 1 s, up to $t =
60$ s: (a) ${\cal R}=0.5$ (b) ${\cal R}=1.0$ (c) ${\cal R}=2.0$.}
\label{new1}
\end{figure}

For ${\cal R}$ of the order of unity, one expects a combination of
these two behaviors. For the assumed exponential intrinsic light
curve, which (for ${\cal {R}}\sim 1$) dies more quickly than the
light curve shape induced by the curvature effect, we expect an
initial phase when the HIC obeys the intrinsic form ($\eta \sim
\gint$) followed eventually by the relation expected from the
curvature effect ($\eta=2$). This transition can be seen in Figure
\ref{new1} for the three cases with ${\cal R} =0.5, 1$ and 2, from
the top to the bottom panels. In each panel we show total spectra for four
different observed times $t$ (solid curves). The circles show the
peak flux and the photon energy at the peak ({\it i.e.} the
observed $\Ep$) for these four and several more total spectra.
This is what the observed HIC will look like. The light dotted
lines show the assumed intrinsic, $\gint=1$, and the curvature-induced,
$\eta=2$, HIC relations. Each one of the total spectra is
a sum of the spectra from different angles $0<\theta<\theta_{\rm
max}(t)$ with different $\Ep$'s having suffered different boosts.
The peak flux and the peak energy of spectra from several angles
are shown by the crosses for each of the four observer times. The
crosses furthest to the left represent the $\theta=0$ spectrum which is
identical to the intrinsic spectrum (of course boosted by
$2\Gamma^2$) and, therefore, lies on the dotted line with slope
$\gint=1$. As we move to the right we see spectra from
successively larger $\theta$s (with varying boosts) till the
crosses furthest to the right which are for $\theta_{\rm max}$,
{\it i.e.}, the radiation emitted at $t=0$, with the flux and
$\Ep$ boosted according the
relations described in the previous section, and lies on the
dotted line with the slope $\eta=2$.

For the lower panel, ${\cal R}=2$, the most intense radiation
comes from $\theta_{\rm max}$. Consequently, the total spectra
obey the large ${\cal R}$, or delta function, HIC relation
($\eta=2$), except for a relatively short but noticeable initial
phase. For smaller ${\cal R}$ this initial phase increases in
length. As evident from the relative position of the crosses in
the panel for ${\cal R}=0.5$, for the first two times shown, the
emission is dominated by the radiation from $\theta=0$ and we
obtain a well defined HIC with $\eta=\gint=1$. But by the time
corresponding to the fourth total spectrum the emission from
$\theta_{\rm max}$ dominates and the HIC reverts to the $\eta=2$
case. The location of the crosses and the final HIC relation for
the ${\cal R}=1$ case is intermediate between the above two cases.

\begin{figure}[]
\epsscale{0.5}
 \plotone{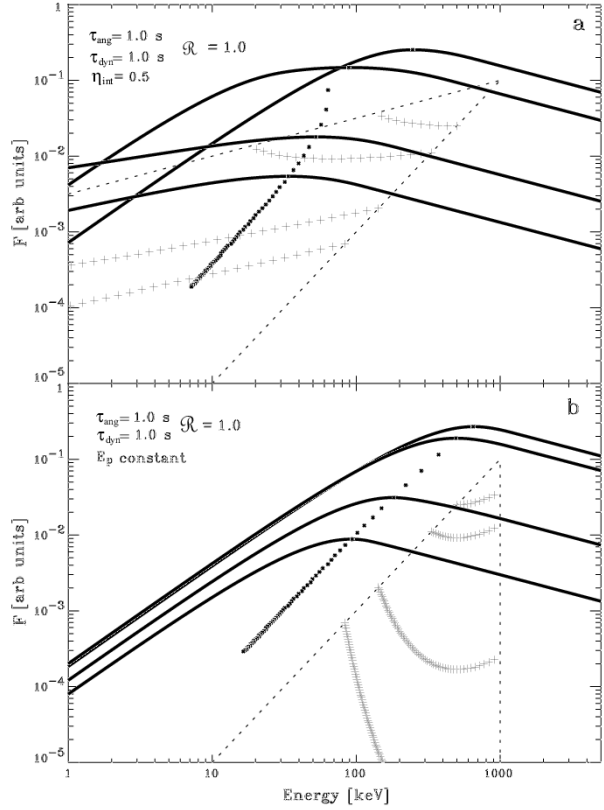}
\caption{Same as Fig. \ref{new1} for ${\cal R}=1.0$,  but two
different intrinsic HIC relations (a) $\eta_{int}= 0.5$ and (b)
$\eta_{int}= \infty$ corresponding to a constant $\Ep'(t)$. The
softening of the observer frame low-energy spectra is clearly seen
in (a) while (b) demonstrates that for steep intrinsic HICs
($\gint
>2$), the slope of the observed HIC will mainly be 2, independent
of the intrinsic slope (see also Fig. \ref{new3}a).
  } \label{new2}
\end{figure}

\begin{figure}[]
\epsscale{0.5}
 \plotone{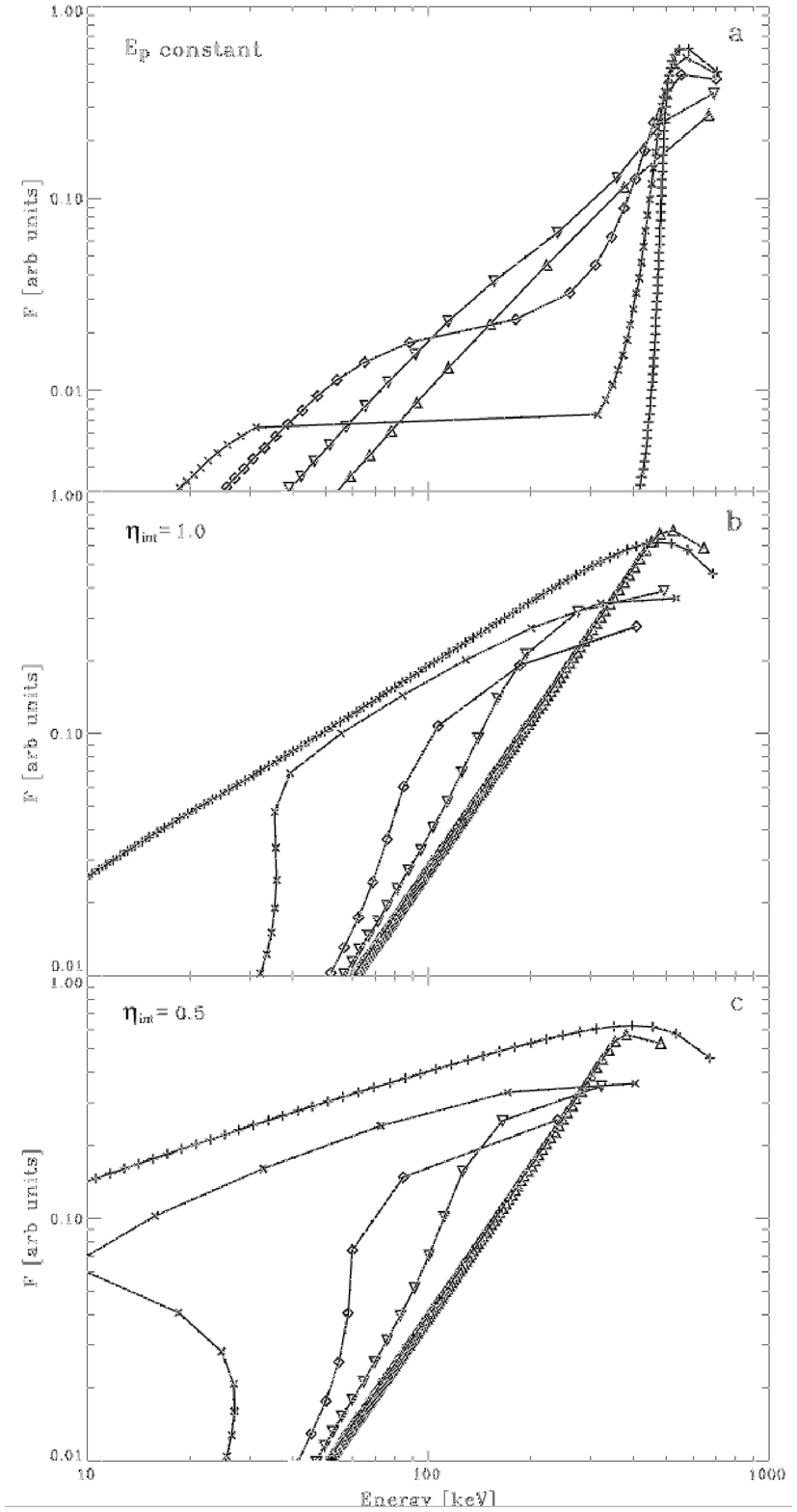}
\caption{Observer frame HIC relations, for different values of
$\R$, produced by three different intrinsic HIC relations: (a)
$\eta_{int}= \infty$ (b) $\eta_{int} = 1$   (c) $\eta_{int}=0.5$.
In panel (a)  the time steps are 2 s  and  ${\cal R}= 0.05$
(crosses), 0.1 (x-es), 0.2 (diamonds), 0.5 (inverted triangles),
1.0 (triangles), while in panel (b) and (c) the time steps are 1 s
and ${\cal R}= 0.05$ (crosses), 0.5 (x-es), 1.0 (diamonds), 2.0
(inverted triangles), 20 (triangles). Note the presence of various
concave HICs which can explain  observed bursts with index $
> 2$, when fitted with a single power law.
  } \label{new3}
\end{figure}

In order to further demonstrate the relative effects  of the
intrinsic and curvature-induced HIC relations, in Figure
\ref{new2} we show two cases with
 ${\cal R}=1$ but with intrinsic power law indexes, $\gint=0.5$ and
$\gint=\infty$. The latter corresponds to a constant $E_{\rm p}$
independent of the flux.  In Figure \ref{new3} we show only the
resultant HIC relation for five different values of ${\cal R}$ for
the above three intrinsic HIC indexes $\gint=.5,1$ and $\infty$.

Several conclusions can be drawn from these figures. The first is
that the observed evolution of $\Ep$ may have nothing to do with
the intrinsic evolution of the emission process, so that care is
necessary in the interpretation of the observed spectral
parameters. This is most clearly evident in Figure \ref{new2}b
where a non-evolving spectrum appears to have a HIC index of $\eta
=2$. It is also clear (from all of the figures) that this most
commonly observed HIC index does not require a very large value of
the ratio $\R$. It is established after a few dynamic times for
$\R=1$ and is omnipresent for $\R>2$, independent of the value of
$\gint$. On the other hand, the intrinsic HIC is essentially what is
observed (at least when the flux is high, say down to
$1$ \% of peak flux) for $\R< 0.5$.

It is clear that if there is some dispersion in the intrinsic
power-law index, $\gint$,  for instance, if it has  a broad,
uniform distribution between 0 and $\infty$, then the observed
distribution will be peaked around $\eta=2$. If the intrinsic
source of radiation is synchrotron emission the preferred value of $\gint$
will lie between $1$ and $2$ \citep{LP02} and the observed
distribution is expected to be even more peaked around 2. However,
a variety of other possible values of this index and deviations
from a simple power law will also be present. This is true
especially for values of ${\cal R} \sim 1$. In addition,
limitations imposed by the instrumentation make it  possible
to follow only a portion of the HIC. This is because the
signal-to-noise of the observations limits the dynamic range of
the observable flux and the finite band width of the instrument
introduces bias against spectra with $\Ep$ outside the band. As
is evident from Figures \ref{new1}, \ref{new2}, \ref{new3} some of
the curves do not follow a perfect power law but are concave, which
is even more pronounced for smaller $\gint$. Thus a simple power
law fit to these curves, excluding the few early phase points
which may appear as part of the rise phase of the pulse, will
yield a steep HIC with $\eta > 2$. In fact inspection of the
pulses with large measured $\eta$-values in the BR01 sample and
illustrated in Figure \ref{fig:moreHICs} show some resemblance to
these concave curves.

The second conclusion is that because the observed spectrum is a
superposition of many intrinsic spectra, it will necessarily be
broader than the intrinsic spectrum and its spectral parameters
could have complex relations with the intrinsic ones. Two obvious
effects of this broadening can be seen in the above figures. One
of these is that the value of $\Ep$ is less well determined
because of the flat tops. This is the cause of the sharp
transition from a high (intrinsic) to a low (boosted) value of
$\Ep$ for the cases with $\R<1$, and $\gint=\infty$. The second
effect of this is that for fits to a finite spectral range of
observations the slope of the flat portion will be identified
mostly with the low energy index and result in $\alpha =-1$
instead of assumed $\alpha = - 2/3$. In some cases this portion
could be identified with high energy index and yield $\beta=-1$
instead of $-2.5$. This effect is more pronounced for smaller
values of $\gint$ (compare Fig. \ref{new2}a and Fig. \ref{new1}b)
because $\Ep$ undergoes a larger variation for small changes in
the flux. More than half of all pulses observed by BATSE have
spectra that do change their shapes in time, often showing a low
energy softening \citep{crider}. The above figures show that such
behavior can be produced purely  by the curvature effect {\it
without} the intrinsic spectrum changing.

\subsection{Effects of Different Intrinsic Light Curves and Rise Times}

\begin{figure}[]
\epsscale{0.5}
 \plotone{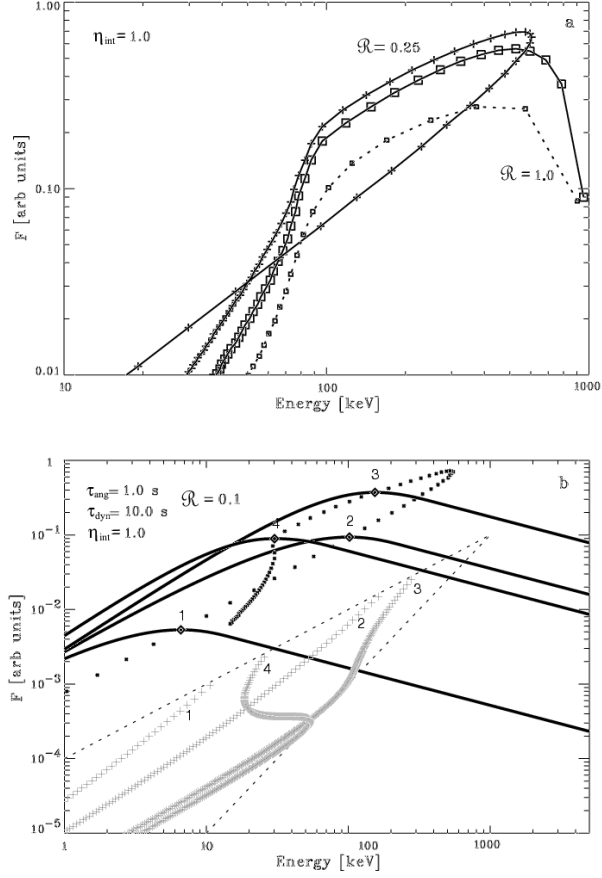}
\caption{ Variations in the spectral behavior due to the variation
in the intrinsic light curve. (a) Same as figure \ref{new3} but
for a Gaussian intrinsic light curve and ${\R}=0.25$, a one-sided,
pure decay Gaussian produces a so-called {\it hard-to-soft} pulse
(squares), while a symmetric Gaussian (crosses) produces a
so-called {\it tracking} (hard-soft-hard) pulse. For the latter
case the rise phase will follow the intrinsic HIC. For comparison,
the dashed line shows the HIC for a one-sided exponential decay
with ${\R}=1.0$ from Fig. \ref{new3}b.
  (b) Same as Fig. \ref{new1} but
for a Gaussian intrinsic light curve with equal rise and decay
time scales. Circles show the HIC relation and the large crosses
show the peaks of the contributing spectra. Times 1 and 2 are
during the rise phase while 3 and 4 are for the decay phase.
 } \label{new45}
\end{figure}

In the discussion above we have used a simple prescription of the
intrinsic light curve with only a decay phase: $\exp(-t/\td)$. Its
actual shape is revealed to the observer only if ${\cal R} \gtrsim
0.5$. Otherwise any claim about the pulse shape is pure
speculation. We have also explored other shapes and found that the
resulting HICs show the same qualitative behavior, albeit that a given
behavior is found for a different value of the characteristic time
scale, $\td$. These similarities and differences can be seen by
comparing the relations marked by squares in Fig. \ref{new45}a,
derived for a pulse with a Gaussian decay phase [$f(t) \propto
\exp[-(t/\td)^2]$, $t > 0$, solid line], and for an exponential
decay [dashed line; compare Fig. \ref{new2}a].

A second aspect of the intrinsic light curve which must be
explored is its finite rise time when this time is not much
shorter than $\td$ and/or $\ta$. The curvature effect during the
rise phase of the intrinsic pulse is somewhat different from its
effect during the decay phase described above. Figure \ref{new45}b
shows this for an intrinsic, complete, not one-sided, Gaussian
light curve. During the intrinsic rise phase (points 1 and 2 in
the figure) the flux will always be dominated by the emission from
$\theta=0$. This is because (i) larger angles sample earlier times
in the intrinsic light curve (which are weaker) and (ii) the boost
factor becomes smaller at larger angles. Therefore, at the early
time of the observed rise phase (before the intrinsic decay has
started) the HIC will simply follow the intrinsic $\gint$, largely
independent of ${\cal R}$. During later times, when the intrinsic
decay starts to be visible at the smallest angles, the behavior
outlined in the previous section will occur. The rise phase will
only affect the observed HIC in the early part of the rise phase.
This is also shown in figure \ref{new45}a where the HICs are
plotted for a one-sided Gaussian with only a decay phase (squares,
solid line) and and a symmetric Gaussian (crosses). Including the
rise phase will produce an up and down (soft-hard-soft) or
so-called  {\it tracking} pulse while the pure decay will produce
a {\it hard-to-soft} evolution, discussed above and by
\citet{ford95}.

In summary, for the intrinsic rise phase the flux will be
dominated by $\theta=0$ and the HIC will follow the intrinsic
$\gint$. This makes the rise phase of a tracking pulse of interest
as it reveals the intrinsic HIC. For the decay phase the ratio
${\cal R}$ determines the behavior.  For ${\cal R} \gg 1$ the flux
from $\theta_{\rm max}$ dominates and $\eta =2$, while for ${\cal
R} \ll 1$ the flux is dominated by $\theta =0$ and the HIC follows
$\gint$. For cases in between, a HIC with $\eta=2$ dominates, for
most of the observable part of the pulse, down to ${\cal R} \sim $
0.5, the exact value depending on the intrinsic light curve shape.
In general the decay phase HIC follows initially a concave curve
sometimes resembling an  S-shaped curve. The transition in these
cases is similar but much broader than those referred to as {\it
track jumps} by BR01 (the case with two parallel HIC curves displaced
from each other), which was explained as two separate spectral
components coming from two overlapping pulses. The S-shape
feature, with a transition from $\eta = \gint$ to $\eta = 2$, can
explain the more moderate changes seen in Figure
\ref{fig:moreHICs}, and demonstrates that the HIC shape can be
used as a diagnostic for the value of ${\cal R}$ and the intrinsic
light curve.

\section{MODIFICATIONS OF THE BASIC MODEL}

\label{sec:caveats}

In deriving the above equations and the HIC we have made several
simplifying assumptions.  We therefore briefly discuss how the
results are affected when we relax some of these restrictions.

\subsection{Spatially Broad Emitting Region}

A limitation of the results presented in \S 3 is the delta
function representation of the spatial distribution of the
emission. The actual emitting region most likely has a finite
width, less or equal to the width  $\Delta'$ of the shell. Here we
assume it to be equal to the shell width, with $j' \propto g(R',
{\Delta'})$ instead of the delta function description used in \S
4. The above approximation is valid for ${\Delta'}/R \ll 1$. This
can be demonstrated assuming a Gaussian distribution,
$g(R',{\Delta'})= 2/({\Delta'} \sqrt{\pi}) \exp \left[
-((R'-R)/{\Delta'})^2\right]$. If we replace the $\delta(R'-R)$
factor in equation (\ref{emiss}) by this form and carry out the
derivation again, then equation (\ref{ekv:F}) becomes
\begin{equation}
\Fb(t)=F_{0} \mu \frac{{\cal{D}}^2(\mu)}{\Gamma^2}\left(
1+\frac{{\Delta'}}{\sqrt{\pi}R} \right)
\end{equation}
If the relation between $\Delta'$ and $R$ is the one given by
equation (\ref{equ:delta}) this becomes $ F(t) \sim F_0\mu
({\cal{D}}/\Gamma)^2 (1 + 1/\Gamma)$, which shows that the width
of the spatial distribution affects the flux (negligible for
$\Gamma \gg 1$) but has no effect at all on the spectrum of the
emission.

\subsection{Geometrical Effects}

Another tacit assumption has been that the emission surface is
spherical, either forming a complete sphere  or, if it is
confined into a jet of opening angle $\theta_{\rm jet}$, that the
LOS intersects the jet at an angle more than $1/\Gamma$ from the
edge of the jet. If the LOS is close to the edge of the jet then
 an extra steepening of the HIC can arise, apart from the
nominal value for the angular spreading case ($\eta = 2$). Let us
consider this effect for the delta function emission, or
${\cal{R}} \rightarrow \infty$. During the early stages of the
pulse an observer will see full $(2\pi )$ annuli with $\Ep$ and
$F$ varying as described in \S 4. However, when the azimuthal
symmetry is broken, {\it i.e.} the annuli are no longer whole, the
flux integrated over the partial annuli will be less compared to
what is expected from a full annulus. However, the observed value
of the $\Ep$ will not be effected as the spectrum will not change.
This will lead to a break in the HIC, from $\eta=2$ to a larger
$\eta$. A corresponding behavior will also be seen in the light
curve. This could explain the most extreme cases in the  BR01
sample.

\subsection{Inhomogeneities}

In the description of the relativistic outflow we have assumed
that the angular distribution of the Lorentz factor is
homogeneous, {\it i.e.}, it is constant over the angles $\theta$
and $\phi$ and that at angles larger than $\theta_{\rm   jet}$,
$\Gamma$ drops abruptly to zero. However, this is just a
simplistic description and does not necessarily describe the
actual situation. Deviations are most probably important at the
edges of the jet. Therefore, if the beaming angle is of the same
size as the jet opening angle and/or if the LOS is close to its
edge, then the intensity will be affected. A variable Lorentz
factor will give rise to a slight deviation from spherical
symmetric outflow. This will  affect only the shape of the light
curve and the time evolution of the observed $\Ep$, but not the
HIC because both $F$ and $\Ep$ will be affected same way.

\subsection{Spectral Variations}

We have also not considered the effects of the intrinsic, rest
frame, spectral-shape variation (changes in $\alpha$ and $\beta$)
during the pulse, but have rather concentrated on the exploration
of the variation of $\Ep$ and flux. As shown above, there is a
softening that occurs naturally in this scenario and hence could
be one of the reasons for the observed softening without the
intrinsic spectrum necessarily changing. For a more general study,
individual cases need to be examined.  This emphasizes again that
the interpretation of the observed spectra and their relation to the
source spectrum, and consequently the emission mechanism, is not
straightforward. This aspect of the problem is, however, beyond
the scope of this paper, but will be dealt with in a future paper.

\section{SUMMARY AND DISCUSSION}
\label{sec:disc}

 We have examined the effect of differences in light travel time due
to the curvature of the expanding shell and determined to what
extent it can affect the width and shape of pulses and their
spectral time evolution. The energy flux $F$ and the photon
energies $E$ will be affected by the angle-dependent Lorentz-boost
 factor $\cal{D(\mu)}$; $F \propto {\cal{D}}^2$ and $E
\propto {\cal{D}}$. The peak energy of the $\nu F_\nu$ spectra,
$\Ep$, will thus follow a hardness-intensity correlation (HIC) $F
\propto E^\eta$ with $\eta=2$. Furthermore, the decay phase of a
pulse will follow the form $(1+t/\ta)^2$. We  show that this
effect should be important for a reasonable choice of parameters
(Lorentz factor, burst energy, shell width etc.) and that these
characteristics agree with the average behaviors found in pulses.

However, the curvature effect can not alone explain the large
observed dispersion of the HIC index $\eta$ . Cases with $\eta$
largely different from $2$ we believe are produced by a finite
dynamic time, $\td \neq 0$. The resulting spectral/temporal
behavior depends mainly on the ratio $\R = \ta/\td$. The intrinsic
HIC (assumed to be a power law with an intrinsic index $\gint$)
will be revealed when $\R < 0.5$ while the behavior expected from
the curvature effect with a HIC index $\eta=2 $ will dominate for
$\R > 2$. For intermediate $\R$s the $ F-\Ep$  relation will
deviate  from a pure  power-law, having a more concave shape (in a
$\log - \log$ plot). A general softening of the spectra with time,
which has been observed, is also expected, independent of any
changes in the intrinsic spectrum, and therefore independent of
the physical environment where the pulses are produced.

An important conclusion of this work is that one must be very
careful in the interpretation of the observed light curves and
spectra, their parameters and evolution. This is because we have
shown that the observed light curve will in most cases be
different from the intrinsic one and the observed spectra will
have a complex relation to the intrinsic ones. The spectra in the
observer frame will be broader and, for instance, the low-energy
power-law slopes $\alpha$ will be softer than the intrinsic ones.
In some cases, flat-topped spectra are produced which, in the
observer frame, appear to have either $\alpha$ or $\beta= -1$.
Furthermore, we also explain the occurrence of pulses whose $\Ep$
track the flux up, and down, during the rise and decay phase,
respectively, as well as the occurrence  of  pulses where the
hardness declines monotonically independent of the rise and fall
of the flux.

Ultimately, we wish to determine the characteristics of the {\it
intrinsic} emission, namely $F'(t)$, $\Ep'(t)$, $\gint$, and if
possible the spectral power law slopes $\alpha$ and $\beta$. In
addition, we want to determine the distance $R$ from the
progenitor where the fireball emits the $\gamma$-rays and to
discern something about the shell width $\Delta'$ and/or its
spreading. Fits to the HIC and observed light curve will be able
to reveal $\R$ and $F'(t)$. Below we discuss the {\it principle}
diagnostics that can be made for three different situations. The
value of the bulk Lorentz factor $\Gamma$ remains an unknown
parameter.

Case I. The observed HIC is a pure $\eta=2$ power law ({\it e.g.}
pulses in Fig.\ref{fig:2cases}): According to our model the
curvature effect is dominant (It could, however, also be due to an
intrinsic $\eta=2$ HIC). The observed light curve will
(asymptotically) follow equation (\ref{ekv:flux}) with $d=2$, from
which one can determine the value of $\ta$. This time constant
determines the distance at which the shell lights up:
\begin{equation}
R = 2 \ta \Gamma^2v=6 \times 10^{14} {\rm cm} \left(\frac{\ta}{1
\rm {s}} \right) \left(\frac{\Gamma}{10^2} \right)^2 \beta
\end{equation}
\noindent
 The observed light curve can, in principle, be
deconvolved,  with equation (\ref{Ft_rel}) as the impulse
response, to obtain $F'(t)$. A better knowledge of $F'(t)$, then
gives a more accurate value of $\R$ (and thereby $\td$) from a fit
to the HIC. Knowing $\td$ one can put constraints on
$\Delta'/{\beta_{\rm sh}}'$ or ${\beta_{\rm sh}}' (\Gamma_{\rm
s}/\Gamma$). Furthermore, the observed, instantaneous spectra will
be results of integrations of the intrinsic spectra along a
$\eta=\gint$ power law. Combining this knowledge with the
observations could reveal $\Ep'(t)$ and $\gint \sim \alpha_{\rm
obs}$ and possibly $\alpha_{\rm int}$.

Case II. The HIC is a pure power law with index $\eta$,
substantially different from $2.0$ (e.g. pulse in Fig.
\ref{fig:moreHICs}d): Here, $\eta=\gint$ and the light curve
should reflect the intrinsic $F'(t)$ (smoothed somewhat by the
curvature effect). The energy evolution follows $\Ep'(t)$. Using
$F'(t)$ a more thorough fit of the HIC can be made giving $\R$,
which gives an estimate of $R {\beta_{\rm sh}}'/ \Delta'$ or
${\beta_{\rm sh}}' (\Gamma_{\rm s}/\Gamma)$ (independent of $R$).
The spectra arise from integrations along $\eta=2$ which,
depending on the details of the case, maybe provide a possibility to
determine $\alpha_{\rm int}$.

Case III. Intermediate cases where S-curves are seen (e.g. pulses
in Figs. \ref{fig:moreHICs}a, b, and c): The low energy section of
the HIC ({\it i.e.} at late times) will follow $\eta=2$ and gives
the value of $\ta$ and $R$. With this knowledge the light curve
can be deconvolved and $F'(t)$ can be found. A fit to the HIC can now
be made to find  $\R$ [which gives $\td$ and $\Delta'$] and $\gint$ which
will be revealed from the early part, [which will allow the
determination of $E'(t)$]. A corresponding softening of the spectra as
described in the paper should be present.

BR01 found that in several GRBs, with two separable pulses, the HIC
index varied less from pulse to pulse in a single burst as
compared to its variation in different bursts. This requires that
the pulses in multi-pulse bursts be produced in shocks created in
a similar environment, with similar values of  $R$, $\Gamma_{\rm
rel}$, $\Gamma$, ${\R}$, $\gint$, $n$ and $B$. This could happen
in a scenario in which the two long pulses are created as two
similar shells catch up with a leading, slower, more bulky shell
that has already been significantly decelerated due to interaction
with the circumburst environment. Such pulses then occur
approximately within the same environment, at roughly the same
distance $R$ (therefore same $\ta$) and $\Gamma_{\rm rel}$. This
scenario also increases the value of $\Gamma_{\rm rel}$, which
implies a higher magnetic field, radiative efficiency, and a minimum
electron Lorentz factor, and a higher synchrotron peak frequency:
\begin{equation}
h \nu_{\rm   s} = \frac{3 e}{4 \pi m_{\rm   e} c} \Gamma B \gamma
_{\rm   e}^2= 5 {\rm   eV} (\Gamma_{\rm
rel}-1)^{2.5}\left(\frac{R}{10^{15}{\rm   cm}} \right)^{-1},
\end{equation}
where we have used the relations for B and $\gamma_{\rm e}$
described at the beginning of \S 3. With $\Gamma_{\rm rel} \sim
100$, $h \nu_{\rm s} = 500$ keV, so that the expected synchrotron
spectrum will peak in the BATSE window and will require no
additional boost, for instance, from Compton upscattering  as in
the Synchrotron-Self-Compton model (SSC) \citep{pan00}. This
scenario is similar to  that of the external shock model normally
proposed for the generation of the afterglows, which has difficulty
to explain the prompt gamma-ray emission because of its high
variability \citep{Fen}. However, the GRBs discussed in this paper
are smooth with few pulses and do  not exhibit the high
variability of more complex bursts, so that this objection is not
applicable.

\acknowledgments
We are grateful to S. Kobayashi, M. Sikora, and L. Borgonovo for
interesting discussions and the referee, J. Norris, for
suggestions that led to an improvement of the paper. F.R. acknowledges
financial support from the Swedish Foundation for International
Cooperation in Research and Higher Education (STINT) and the
Ludovisi Boncompagni, n\'ee Bildt foundation. This research made
use of data obtained through the HEASARC Online Service provided
by NASA's Goddard Space Flight Center.


\newpage

\begin{thebibliography}{}

\bibitem[Band et~al. (1993)]{band} Band, D., et~al.\ 1993, \apj,  413, 281

\bibitem[Borgonovo \& Ryde (2001)]{BR01} Borgonovo, L., \& Ryde, F. 2001,
        \apj, 548, 770

\bibitem[Borgonovo et al. (2002)]{BRVS} Borgonovo, L., Ryde, F., de Val Borro,
M., \& Svensson, R. 2002,  in the proceedings of
        'Gamma-Ray Burst and Afterglow Astronomy 2001', Woods Hole,
        MA, in press

\bibitem[Crider et al. (1997)]{crider} Crider, A., et al. 1997, ApJ, 479, L39

\bibitem[Eriksen \& {Gr{\o}n} (2000)]{eri00} Eriksen, E., \& Gr{\o}n, {\O}. 2000,
Amer. J. Phys., 68, 1123

\bibitem[Fenimore et al. (1996)]{Fen}Fenimore, E. E., Madras, C.
        D., \& Nayakshin, S. 1996, ApJ 473, 998

\bibitem[Fenimore \& Sumner (1997)]{FenRik}Fenimore, E., \& Sumner,
        M. C. 1997, All-Sky X-Ray Observations in the Next Decade, 167

\bibitem[Fishman et al. (1994)]{fish94} Fishman, G. J., et~al.\ 1994,
        ApJS, 92, 229

\bibitem[Ford et al. (1995)]{ford95}Ford, L. A., et al. 1995, ApJ, 439, 307

\bibitem[Frail et al. (2001)]{fra01}Frail, D. A., et al. 2001, ApJ,
562, L55

\bibitem[Gissellini et al. (2000)]{ghis}Gisellini, G., Celotti, A., \& Lazzati, D.
        2000, MNRAS, 313, L1

\bibitem[Granot et al. (1999)]{Gra99}Granot, J., Piran, T., Sari,
        R. 1999, ApJ, 513, 679

\bibitem[Golenetskii et al. (1983)]{gol83} Golenetskii, S. V.,
        Mazets, E. P., Aptekar, R. L., \& Ilyinskii, V. N.
        1983, Nature, 306, 451

\bibitem[Kargatis et al. (1995)]{kar95}Kargatis, V. E., et~al.\ 1995,
        A\&SS, 231, 177

\bibitem[Kocevski \& Liang (2001)]{KL} Kocevski, D., \& Liang, E. 2001,
        in AIP Conf. Proc. 586, Relativistic Astrophysics: 20th Texas
        Symposium, ed. J. C. Wheeler, \& H. Martell
        (New York: AIP), 623

\bibitem[Lee et al. (2000a)]{lee1}Lee A., Bloom, E. D., \& Petrosian, V. 2000a, ApJS, 131, 1

\bibitem[Lee et al. (2000b)]{lee2}Lee A., Bloom, E. D., \& Petrosian, V. 2000b, ApJS, 131, 21

\bibitem[Liang (1997)]{liang} Liang, E. P. 1997, \apj, 491, L15

\bibitem[Liang \& Kargatis (1996)]{LK96} Liang, E. P., \& Kargatis,
        V. E. 1996, Nature, 381, 495

\bibitem[Lloyd \& Petrosian (2000)]{LP01}Lloyd, N., \& Petrosian, V. 2000,
 ApJ, 543, 722

\bibitem[Lloyd \& Petrosian (2002)]{LP02}Lloyd, N., \& Petrosian, V. 2002,
 ApJ, 565, 182

\bibitem[Lyutikov \& Blackman (2001)]{lyu}Lyutikov, M. \& Blackman, E. G. 2001,
MNRAS 321, 177

\bibitem[Norris et al. (1996)]{norris}Norris, J. P., Nemiroff, R. J., Bonnell, J. T.,
Scargle, J. D., Kouveliotou, C., Paciesas, W. S., Meegan, C.A., \&
Fishman, G. J. 1996, \apj, 459, 393

\bibitem[Norris, Marani, \& Bonnell (2000)]{norris2}Norris, J. P.,
Marani, G. F.  Bonnell, J. T. 2000, \apj, 534, 248

\bibitem[Panaitescu \& M\'esz\'aros (2000)]{pan00} Panaitescu, A.,
        \& M\'esz\'aros, P. 2000, \apj, 544, L17

\bibitem[Rees \& M\'esz\'aros (1992)]{ree92} Rees, M. J., \& M\'esz\'aros, P.
        1992, MNRAS, 258, 41

\bibitem[Rybicki \& Lightman (1979)]{RL} Rybicki, G. B., \& Lightman,
        A. P. 1979, Radiative Processes in Astrophysics (New York: Wiley)

\bibitem[Ryde, Borgonovo, \& Svensson (2000)]{RBS00} Ryde, F.,
        Borgonovo, L., \& Svensson, R. 2000,
        in AIP Conf. Proc. 526, Gamma-Ray Bursts, 5th Huntsville
        Symposium, ed. R. M. Kippen, R. S. Mallozzi, \& G. J. Fishman
        (New York: AIP), 180

\bibitem[Ryde, Kocevski \& Liang (2002)]{RKL} Ryde, F., Kocevski, D.,
\& Liang, E., in the proceedings of 'Gamma-Ray Burst and Afterglow
Astronomy 2001', Woods Hole, MA, in press

\bibitem[Ryde \& Svensson (2000), RS00]{RS00} Ryde, F., \& Svensson, R.
        2000, \apj, 529, L13

\bibitem[Ryde \& Svensson (2002)]{RS02} Ryde, F., \& Svensson, R.
        2002, ApJ, 566, 210

\bibitem[Soderberg \& Fenimore (2001)]{sod01} Soderberg, A. M. ,
\& Fenimore, E. E. 2001, in Gamma-Ray Bursts in the Afterglow Era,
ed. E. Costa, F. Frontera, \& J. Hjorth (Berlin Heidelberg: Springer), 87

\bibitem[Spada, Panaitescu, \& M\'esz\'aros (2000)]{SPM00}
        Spada, M., Panaitescu, A., \& M\'esz\'aros, P. 2000, ApJ 537, 824

\bibitem[Tavani (1996)]{tavani}Tavani, M. 1996, ApJ 466, 768

\end{thebibliography}
\end{document}